\documentclass{physeauth}
\usepackage{graphicx}
\usepackage{amsmath}
\usepackage{amssymb}

\def \beq {\begin{equation}}
\def \eeq {\end{equation}}

\newcommand{\fig}[1]{Fig.~\ref{#1}}

\begin{document}

\setlength\arraycolsep{2pt}

\begin{frontmatter}

\title{Towards Quantum Transport
for Nuclear Reactions}

\author{Pawel Danielewicz\thanksref{thank1}},
\author{Arnau Rios} and
\author{Brent Barker}

\address{National Superconducting Cyclotron Laboratory and Department of Physics and Astronomy, Michigan State University, East Lansing, MI 48824-1321, USA}

\thanks[thank1]{
Corresponding author.
E-mail: danielewicz@nscl.msu.edu}

\begin{abstract}
Nonequilibrium Green's functions represent a promising tool for describing central nuclear reactions.  Even at the single-particle level, though, the Green's functions contain more information that computers may handle in the foreseeable future.  In this study, we investigate whether all the information contained in the Green's functions is necessarily relevant when describing the time evolution of nuclear reactions. For this, we carry out mean-field calculations of slab collisions in one dimension.
\end{abstract}

\end{frontmatter}


\section{Simulations of Nuclear Reactions}
Historically, the description of central nuclear reactions has involved a handful of methods. On the one hand, the time-dependent Hartree-Fock (TDHF) approach has been used to study low-energy reactions~\cite{Bonche1976,negele82,Umar05}. On the other hand, the semiclassical Boltzmann equation (BE) has been applied to the reactions at intermediate and high-energies \cite{bertsch88}. Molecular dynamics approaches, sharing elements of  both TDHF and~BE, have also been employed \cite{aic86,ono92}. Either TDHF or BE have some serious limitations that are principally remedied within the nonequilibrium Green's function method \cite{kadanoff,danielewicz84}, as will be indicated in the following. Even at the basic single-particle level, however, the Green's functions method involves the handling of a~vast amount of information, likely to overwhelm the capabilities of computing systems.  Here, we investigate whether all the information in the Green's functions is equally important for the reaction dynamics.  To this end, we study the dynamics in one dimension in the absence of correlations.  These results are principally equivalent to one-dimensional TDHF calculations, but are performed in terms of Green's functions rather than single-particle wavefunctions.

Within the TDHF method, the system wavefunction~$\Phi$ is approximated in terms of one Slater determinant of single-nucleon wavefunctions, $\lbrace \phi_j \rbrace_{j=1}^A$:
\begin{equation}
\Phi \left( \lbrace {\pmb r}_j \rbrace_{j=1}^A, t \right)
= \frac{1}{A!} \sum_{\sigma} \prod_{k=1}^A (-1)^{\text{sgn} \, \sigma} \phi_k \left( {\pmb r}_{\sigma(k)}, t \right) \, .
\end{equation}
The single-nucleon wavefunctions follow wave equations in terms of a self-consistent mean field $U$:
\begin{equation}
i \frac{\partial}{\partial t} \, \phi_j =
\left( - \frac{\nabla^2}{2m} + U(\lbrace \phi_k \rbrace) \right) \phi_j \, .
\label{eq:TDHF}
\end{equation}

\begin{figure}
  \centerline{\includegraphics[width=.44\textwidth]{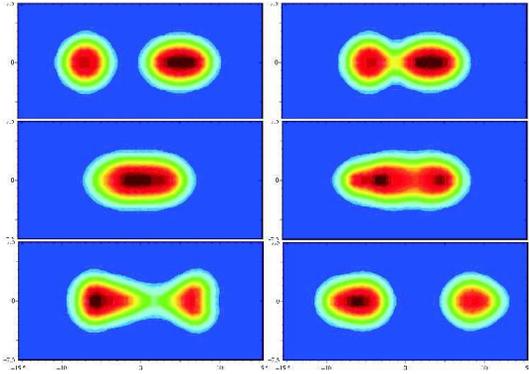}}
  \caption{Contour plots for nucleon density in a head-on collision of $^{16}$O + $^{22}$Ne at $E_\text{cm} = 95 \, \text{MeV}$, from TDHF calculations of Ref.~\cite{Umar05}.}
  \label{fig:TDHF_collision}
\end{figure}

\begin{figure}
  \centerline{\includegraphics[width=.42\textwidth]{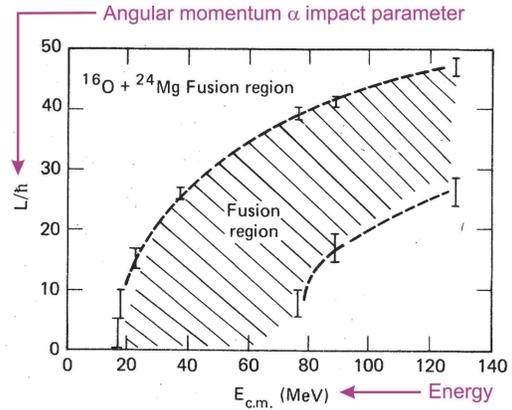}}
  \caption{Fusion region in the plane of angular momentum vs center-of-mass energy for a $^{16}$O + $^{24}$Mg reaction, from the TDHF calculations of Ref.~\cite{krieger81}.  The outer boundary of the region is associated with the fact that nuclei, in their relative motion, must overcome combined Coulomb and centrifugal barriers.}
  \label{fig:TDHF_low_l}
\end{figure}

The first applications of TDHF in the nuclear field have included fusion reactions.  An example of the head-on reaction of $^{16}$O + $^{22}$Ne, at $E_\text{cm} = 95 \, \text{MeV}$, from a calculation by Umar and Oberacker \cite{Umar05}, is shown in Fig.~\ref{fig:TDHF_collision}.  In this particular collision, the nuclei form a compound object that subsequently breaks up, \emph{i.e.} no fusion occurs.  TDHF calculations, in fact, predict a~so-called low-$\ell$ window, regarding fusion reactions, that develops at high collision energies \cite{negele82,krieger81}, see Fig.~\ref{fig:TDHF_low_l}. While nuclei fuse in the more peripheral reactions (at high angular momenta), they fail to do so in the more central reactions (at low angular momenta).  No evidence of such a~low-$\ell$ window has been found experimentally~\cite{szanto81}.  The failure of TDHF with that respect has been attributed to the lack of dissipation, associated to the missing correlations in a mean-field description.  In this particular context, the interest in TDHF as a general theoretical method for addressing central reactions, has seriously waned.

Higher-energy central nuclear reactions have been commonly described \cite{bertsch88,dan91} in terms of the BE for the evolution of the phase-space distribution functions, $f({\pmb r},{\pmb p},t)$, of nucleons and other particles:
\begin{equation}
\frac{\partial f}{\partial t} + \frac{\partial \omega_{\pmb p}}{\partial {\pmb p}} \,
\frac{\partial f}{\partial {\pmb r}}
- \frac{\partial \omega_{\pmb p}}{\partial {\pmb p}}\,
\frac{\partial f}{\partial {\pmb p}} = I\lbrace f \rbrace \, .
\label{eq:BE}
\end{equation}
Here, $\omega_{\pmb p} ({\pmb r},t)$ is the quasiparticle energy for a nucleon with momentum~${\pmb p}$, at location ${\pmb r}$, and $I$ is a collision integral.  BE may be solved following test-particle method, where the phase-space distribution is represented in terms of test-particles at phase-space locations $({\pmb r}_i(t), {\pmb p}_i (t))$,
\begin{equation}
f({\pmb r},{\pmb p},t) \simeq {\mathcal W} \sum_i \delta({\pmb r} - {\pmb r}_i(t)) \,
\delta({\pmb p} - {\pmb p}_i(t)) \, .
\end{equation}
The test-particle locations obey Hamilton-type equations that follow from integrating the l.h.s.\ of Eq.~\eqref{eq:BE},
\begin{equation}
\dot{\pmb r}_i = \frac{\partial \omega_{\pmb p}}{\partial {\pmb p}} \, ,
\hspace*{2em}
\dot{\pmb p}_i = - \frac{\partial \omega_{\pmb p}}{\partial {\pmb r}} \, .
\end{equation}
Moreover, the test-particles undergo collisions which accomplish a
Monte-Carlo integration of the collision integral~$I$~\cite{bertsch88,dan91}.

\begin{figure}
  \centerline{\includegraphics[width=.38\textwidth, height=.30\textheight]{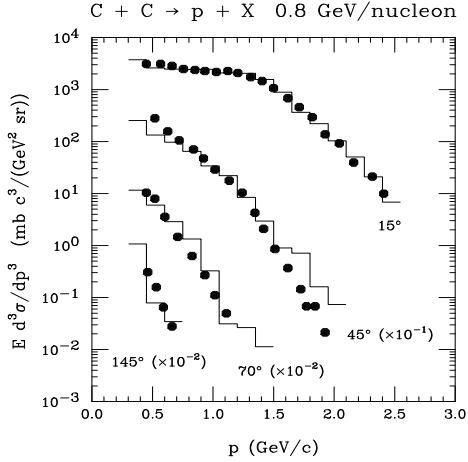}}
  \caption{Proton spectra from an 800 MeV/nucleon $^{12}$C + $^{12}$C reaction.  Dots represent data of Ref.~\cite{nagamiya81} and histograms represent BE calculations of Ref.~\cite{dan91}.}
  \label{fig:CCp}
\end{figure}

BE has been fairly successful in describing many aspects of higher-energy reactions, see e.g.~Fig.~\ref{fig:CCp}.  However, the use of BE in reactions has been criticized on theoretical grounds.  BE relies on the quasiparticle picture and simple estimates \cite{danielewicz84} indicate that particle scattering rates are comparable to particle energies, which undermines that picture.  In addition, in this context it is theoretically difficult~\cite{dickhoff98} to separate collisional effects, described with cross-sections and entering the integral~$I$, from mean-field effects entering the quasiparticle energies, such as in $\omega_{\pmb p} = p^2/2m + U$.

\section{Kadanoff-Baym Equations}

With quantum nonequilibrium many-body theory, the dynamics of a system, starting from an initial state~$|\Phi \rangle$, may be described, in a self-contained manner, in terms of a generalized single-particle Green's function
\begin{equation}
i \, G(1,1') = \langle \Phi | T \left\lbrace \psi (1) \, \psi^\dagger (1')   \right\rbrace    | \Phi \rangle \, .
\end{equation}
Here, $T$ is a generalized time-ordering operator which allows for either ordering of the single-particle operators $\psi$ and $\psi^\dagger$.  The arguments represent space and time, $1 \equiv ({\pmb r}_1, t_1)$ and discrete indices are ignored in the following.  The generalized Green's function satisfies an integral Dyson equation,
\beq
G = G_0 + G_0 \, \Sigma \, G \, ,
\eeq
in terms of the noninteracting Green's function $G_0$ and the self energy $\Sigma$.  The latter is given by
\beq
i \, \Sigma(1,1') = \langle \Phi | T \left\lbrace j (1) \, j^\dagger (1')   \right\rbrace    | \Phi \rangle_\text{irred} \, ,
\eeq
where $j$ is the source for the field $\psi$,
\beq
\left( i \, \frac{\partial}{\partial t_1} + \frac{{\pmb \nabla}_1^2}{2 m} \right)  \psi(1) = j(1) \, .
\eeq

Application of the operator $G_0^{-1}$, inverse to $G_0$ in space-time, to both sides of the Dyson equation, yields an integro-differential equation of motion for the Green's function~$G$.  For the specific order of operators in the expectation value for $G$, such as in $-i G^<(1,1') = \langle \psi^\dagger (1') \, \psi (1)  \rangle$, that equation reduces to the Kadanoff-Baym (KB) equations,
\beq
\begin{split}
\bigg( i \, \frac{\partial}{\partial t_1} & + \frac{{\pmb \nabla}_1^2}{2 m} \bigg) \, G^\lessgtr (1, 1') \\  = & \int \text{d} 1'' \, \Sigma^+ (1, 1'')  \,  G^\lessgtr (1'', 1')\\ & +
\int \text{d} 1'' \, \Sigma^\lessgtr (1, 1'') \, G^- (1'', 1') \, .
\end{split}
\eeq
From the two Green's functions, $G^\lessgtr$, $-i G^<$ yields the 1-particle density matrix when computed at $t_1 = t_{1'}$, i.e.~it can be used to calculate all the 1-body observables.  For interactions of exclusively 2-body type, the function $-i G^<$ can be further used to calculate the net system energy.

A variety of physics phenomena, within different situations, can be described by employing consistent approximations to~$\Sigma$ and the dynamics.  When $\text{Re} \, \Sigma^+ \sim U \gg \Sigma^\lessgtr$, the mean-field (TDHF) approximation applies, for which the Green's function may be approximated by
\beq
-i \, G^<(1,1') \approx \sum_{j=1}^A \phi_j(1) \, \phi_j^*(1') \, .
\label{eq:Gdiag}
\eeq
When scales associated with space-time variation in the average of the Green's function arguments, $(1 + 1')/2$, are much larger than the scales associated with variation in the difference of the arguments, $(1 - 1')$, the quasiparticle approximation applies.  Within that approximation, the Green's function $G^<$ may be represented as
\beq
-i \, G^< (1,1') \approx \int \text{d} {\pmb p} \, f({\pmb p}, 1) \, \text{e}^{i \, {\pmb p} ({\pmb x}_1 - {\pmb x}_{1'}) - i \, \omega_{\pmb p} (t_{1}-t_{1'}) } \, ,
\eeq
and a substitution of such a representation to the KB equations yields the BE.

\begin{figure}
  \centerline{\includegraphics[width=.46\textwidth]{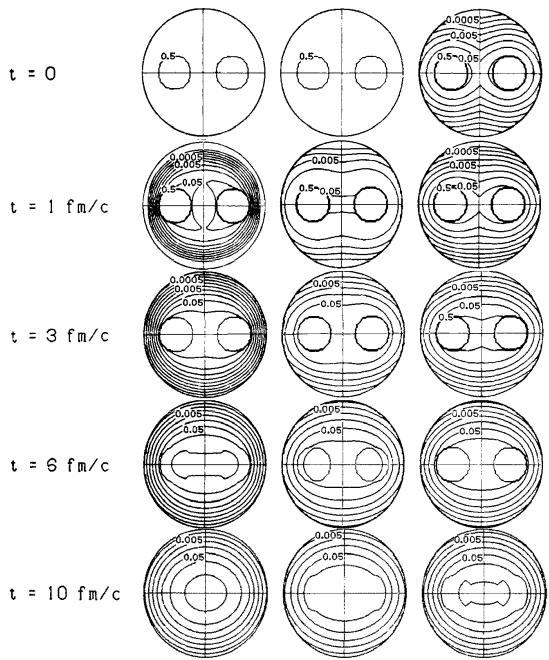}}
  \caption{Contour plots of the distribution of nucleons in momentum, for the evolution of interpenetrating nuclear matters that start, at $t=0$, from two separated Fermi spheres with displacement corresponding to the incident energy of $E_\text{lab} = 400 \, \text{MeV/nucleon}$.  The left column displays the evolution for BE and the center and right columns display the evolution for KB equations when starting, respectively, from an uncorrelated and correlated initial states, after Ref.~\cite{danielewicz84}.}
  \label{fig:equmat}
\end{figure}

Within nuclear physics, the nonequilibrium Green's functions and KB equations have been primarily used for derivations \cite{dan91,ivanov00,leupold00,cassing00}.  In one early case, KB equations have been solved directly for a uniform system of equilibrating interpenetrating nuclear matters \cite{danielewicz84}, see Fig.~\ref{fig:equmat}.  The system of nuclear matters represents the first stages of a central collision of nuclei at intermediate incident energies.
The KB results for this model uniform system have been compared to BE calculations~\cite{danielewicz84,kohler95}.  Differences in these results, see Fig.~\ref{fig:equmat}, in the way in which the momentum space gets populated, and in the pace of approach to equilibrium, demonstrate the importance of going beyond the quasiparticle approximation in the description of intermediate-energy central reactions.  More results from solving directly the KB equations have been obtained in other fields~\cite{haug,kwong98,dahlen07}.

\section{Towards Direct Application of the Green's Functions to Reactions}

The Green's function technique is attractive in that it allows one to incorporate different types of correlations within $\Sigma$ in the system dynamics, without resorting to the quasiparticle limit.  The description of the reaction can be made consistent with the description of the structure of the reacting nuclei~\cite{muther00,dickhoff04}.  The serious disadvantage of the method is that potentially enormous amount of information, representing the 8-dimensional structure of the Green's functions, must be monitored for a reaction.  A secondary deficiency is that potentially two separate calculations need to be carried out to complete a reaction simulation: one preparing the system for the reaction and second advancing the reaction in time.

The 8-dimensional information in the Green's function can be generally understood as mappable onto the 3 components of particle momentum, onto energy (an~independent variable outside of the quasiparticle limit) and onto time and 3 components of position where the particle momentum and energy are considered.  An increase in the dimension by one variable, the energy, as compared to BE, might be considered a minor complication.  However, the serious issue is that no positive-definite quantity can be associated with the Green's functions, such as $f$, thus precluding the possibility of applying the test-particle method.  The advantage of the test-particle method is that it allows one to concentrate on the parts of the phase space that matter at a~given stage of the reaction.  Without this method, a mesh in space-time needs to be established and different space-time regions may potentially need to be monitored, irrespectively of whether or not those regions dominate a~given stage of a reaction.  An alternative to the mesh is projection onto mean-field wavefunctions \cite{tohyama87}, but that can limit excitation energies in the reaction simulations.  In the following, we shall try to assess whether the whole involved space-time region needs indeed to be monitored within the mesh set up for the reaction.  We shall carry out the assessment in one dimension, where the cost of monitoring the whole space-time is moderate, unlike in three dimensions, where the cost of such monitoring is devastating.  We shall also consider the possibility of preparing the initial state when following time evolution such as in a reaction.

\subsection{Slabs in One Dimension}

To assess the importance of different regions in the domain of the Green's functions, we consider interactions of nuclear slabs, for which density changes in the $x$-direction and for which transverse degrees of freedom are frozen.  Those slabs are spin-isospin symmetric and follow mean-field dynamics with Coulomb interactions switched off.  Under these circumstances, the dynamics is consistently described in terms of the Green's function $G^<$ at equal time arguments and at equal transverse positions.  The trace of the density matrix over spin and isospin is then
\beq
\rho(x, x', t) \equiv - 4 \, G^<(x,y,z,t,x',y,z,t) \, ,
\eeq
where the factor of 4 represents the spin-isospin degeneracy.  The density matrix satisfies the equation of motion
\beq
\begin{split}
i \, \frac{\partial}{\partial t} \, \rho (x, x', t) =  &
\left[ - \frac{1}{2m} \frac{\partial^2 }{\partial x^2} + U(x,t)  \right] \, \rho (x, x', t)\\
& - \left[ - \frac{1}{2m} \frac{\partial^2 }{\partial x^{\prime 2}} + U(x',t) \right] \, \rho (x, x', t) \, ,
\end{split}
\label{eq:rhoeom}
\eeq
where $U$ is a Skyrme mean-field of the form
\beq
U = \frac{3}{4} \, t_0 \, n(x,t) + \frac{2+\sigma}{16} \, t_3 \, \left[ n(x,t) \right]^{1+\sigma} \, ,
\eeq
and the density $n$ represents diagonal elements of the density matrix, $n(x,t) = \rho(x,x,t)$.  The parameter constants in the Skyrme interaction are fitted to the properties of nuclei.   If the density matrix is diagonalized in terms of eigenstate wavefunctions, as in \eqref{eq:Gdiag}, then Eq.~\eqref{eq:rhoeom} can be solved by having the wavefunctions follow the TDHF equations \eqref{eq:TDHF}.  However, in the context of the Green's functions, it is the equation for the density matrix,  Eq.~\eqref{eq:rhoeom}, that is of interest.  Net kinetic energy associated with the $x$-direction, per unit transverse area, is
\beq
K(t) =  \frac{\hbar^2}{2m} \int \text{d}x \, \frac{\partial^2}{\partial x \, \partial x'} \left. \rho(x,x',t) \right|_{x'=x} \, ,
\eeq
and the net potential energy per unit area is
\beq
V(t) = \int  \text{d}x \, \left[
\frac{3}{8} t_0 \left[ n(x) \right]^2 + \frac{1}{16} t_3 \left[ n(x) \right]^{2+\sigma} \right] \, .
\eeq

In evolving the slab systems, we use periodic boundary conditions for the calculational region and employ the split-operator method to advance the density matrix in time.  Thus, the change in the density operator over the time-interval $\Delta t$ may be represented as
\beq
\begin{split}
\rho (x, & x', t+ \Delta t)\\ & = \text{e}^{-i({\mathcal K}+U) \, \Delta t} \,
\rho (x,x', t) \, \text{e}^{i({\mathcal K}+U) \, \Delta t} \, ,
\label{eq:rhodt}
\end{split}
\eeq
where ${\mathcal K}$ is the kinetic-energy operator.  Using the Baker-Campbell-Hausdorff formula, the exponentials of a sum of non-commuting operators in \eqref{eq:rhodt} can be rewritten into a product of exponentials of individual operators
\beq
\text{e}^{i({\mathcal K}+U) \, \Delta t} = \text{e}^{iU \frac{\Delta t}{2}} \, \text{e}^{i{\mathcal K} \, \Delta t} \,
\text{e}^{iU \frac{\Delta t}{2}} + {\mathcal O} \left[ (\Delta t)^3 \right] \, ,
\eeq
with the accuracy of the second order in $\Delta t$.
The split-operator method relies on the above rearrangement and the fact that the operators ${\mathcal K}$ and $U$ are diagonal either in the momentum or configuration space.  Application of an exponential in either space amounts then just to a multiplication by a phase factor.  Within the method, the advancement of the density matrix in time can be then accomplished by transforming the matrix between configuration and momentum spacing, exploiting Fast-Fourier-Transformation, and multiplication by the respective phase factors.  Unitarity is built into the method.

\begin{figure}
  \centerline{\includegraphics[width=.41\textwidth,height=.28\textheight]{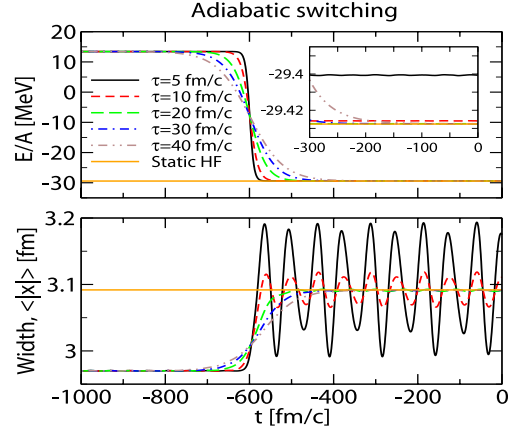}}
  \caption{Evolution of the energy per nucleon (top) and of the size of density distribution (bottom) when adiabatically switching the potential from a harmonic oscillator to a mean-field one, for different values of the $\tau$ parameter in Eqs.~\eqref{eq:Uf} and \eqref{eq:ft=}.}
  \label{fig:ad_ener}
\end{figure}

With regard to the evolution, we first consider the possibility of preparing the initial state for a reaction in an adiabatic fashion \cite{messiah}.  We start out with a state obtained by filling out two lowest orbits of the harmonic oscillator potential and we gradually change the potential from that of the harmonic oscillator to the mean-field one, with
\beq
U_f(x,t) = f(t) \, \frac{k}{2} x^2 + [1 -f(t)] \, U(n(x,t)) \, ,
\label{eq:Uf}
\eeq
where we typically use
\beq
f(t) = \frac{1}{1 + \text{e}^{(t - \tau_0)/\tau}} \, .
\label{eq:ft=}
\eeq
Figure \ref{fig:ad_ener} shows the evolution of the net energy per nucleon and net size of the density distribution, when gradually changing the potential.  Notably, the energy itself is not that good indicator of the arrival at the ground state, because it depends quadratically on the deviation from the ground state.  Characteristics of the density, on the other hand, depend linearly on the deviation and are better indicators.  Values of $\tau \gtrsim 20 \, \text{fm/}c$ lead to a very good approximation of the mean-field ground-state.  

\begin{figure}
  \centerline{\includegraphics[width=.48\textwidth]{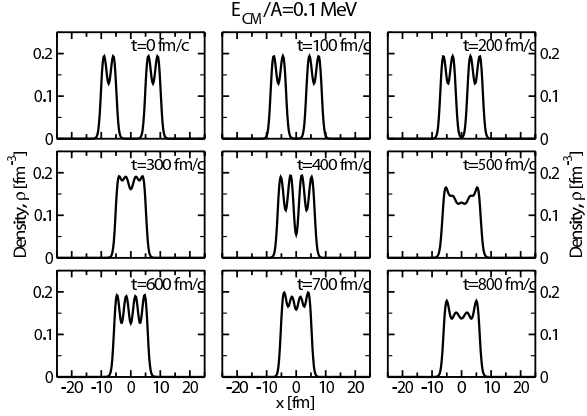}}
  \caption{Evolution of density for slabs colliding at c.m.\ energy of 0.1~MeV/nucleon.}
  \label{fig:den01}
\end{figure}

\begin{figure}
  \centerline{\includegraphics[width=.48\textwidth]{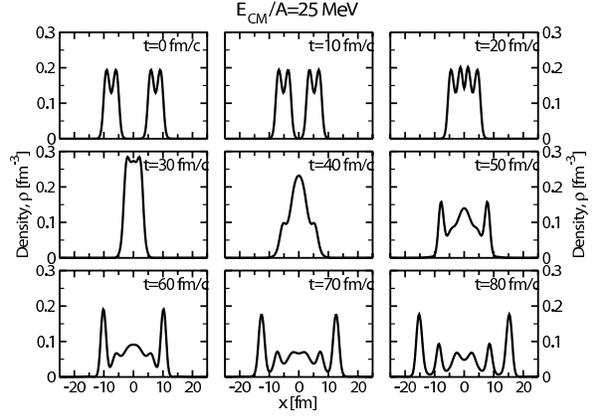}}
  \caption{Evolution of density for slabs colliding at c.m.\ energy of 25~MeV/nucleon.}
  \label{fig:den25}
\end{figure}

We next consider slab collisions.  For the collisions, the slabs are boosted and positioned against each other.  In a boost, the individual density matrices transform according to
\beq
\rho (x,x',t=0) \rightarrow
\text{e}^{ip_\text{cm} x} \, \rho (x,x',t=0) \, \text{e}^{-ip_\text{cm}x'} \, ,
\eeq
where $p_\text{cm}$ is the initial c.m.\ momentum per nucleon.  Figure \ref{fig:den01} shows the evolution of a slab reaction at incident c.m.\ energy of 0.1~MeV/nucleon.  In the absence of Coulomb interactions, fusion occurs at this energy.  At energies $15 \, \text{MeV/nucleon} \gtrsim E_\text{cm} \gtrsim 0.6 \, \text{MeV/nucleon}$, the slabs pass through each other, but get excited and may break up.  At still higher energies, $E_\text{cm} \gtrsim 15 \, \text{MeV/nucleon}$, the identity of the original slabs is largely lost, see Fig.~\ref{fig:den25}, and the system breaks into a multitude of fragments.  In the nuclear reaction terminology, this is called multifragmentation.

\begin{figure}
  \centerline{\includegraphics[width=.48\textwidth]{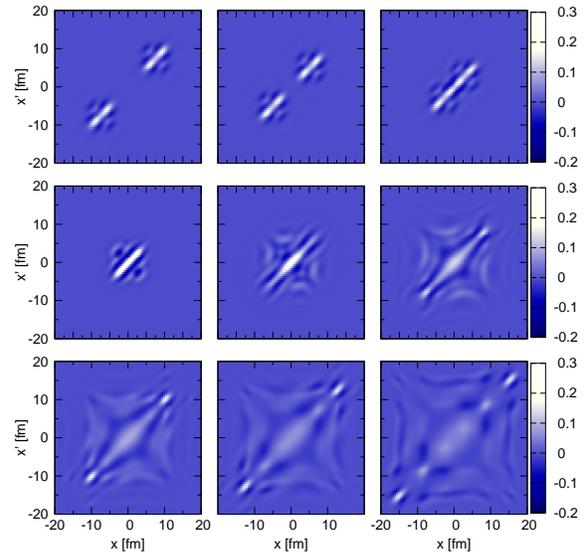}}
  \caption{Intensity plots for the real part of the density matrix $\rho(x,x',t)$ in the collision of slabs at $E_\text{cm} = 25 \, \text{MeV/nucleon}$. The snapshots are taken at the same times as in \fig{fig:den25}}
  \label{fig:gxx}
\end{figure}

Figure \ref{fig:gxx} shows next the intensity plot for the real part of the density matrix for the system of slabs colliding at the c.m.\ energy of 25~MeV/nucleon.  The density matrix is always largest along the $x=x'$ diagonal and the real components tend to marginally dominate over the imaginary components.  The values of the matrix along the diagonal represent the local density, i.e.~Fig.~\ref{fig:den25}.  Early on in the reaction, the more significant values of the density matrix are limited to square-like regions. The diagonals of these squares are collinear with the diagonal of the density matrix and their size represents the support of the nucleon wavefunctions within the individual slabs.  When the system fuses, the size of this region expands to one representing the size of the fused system.  The system evolves further and breaks up into pieces, and the region of significant values for the density matrix expands.  Besides regions adjacent to the diagonal, islands and streaks of significant values emerge, moving away from the diagonal with time.  These are associated with the fragmentation of the original single-nucleon wavefunctions, into pieces taking off with different fragments.  The regions of significant off-diagonal values for $\rho(x,x',t)$ represent phase correlations between the amplitudes for the nucleons, from the individual original states, to move away with one or another fragment.

\begin{figure}
  \centerline{\includegraphics[width=.24\textwidth]{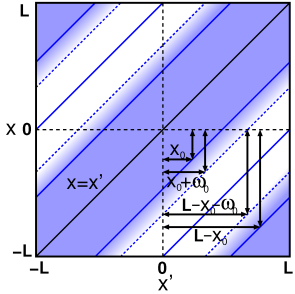}}
  \caption{Manipulation of off-diagonal elements of the density matrix $\rho(x,x',t)$ in testing.  White regions represent elements that get suppressed.}
  \label{fig:planeg}
\end{figure}

The need to monitor far off-diagonal values of the Green's functions would have been devastating for the capability of carrying reaction simulations in 3 dimensions.  However, if the fragments from the break-up are not likely to ever meet again, the phase information in the far off-diagonal values should not matter.  To test the importance of these far off-diagonal elements, we repeat the calculations employing now a strong imaginary superoperator potential that suppresses elements away from the diagonal axis and leaves the vicinity of the axis in the density matrix intact, see Fig.~\ref{fig:planeg}.  Note that the periodic boundary conditions for the system imply a tile-like periodicity of the density matrix $\rho$ in the $x$-$x'$ plane, with the values on the diagonal repeated next on the lines passing through the corners at $(-L,L)$ and $(L,-L)$.  When the periodicity is imposed onto the superoperator, the suppression of the matrix elements has to occur in valleys in the $x$-$x'$ plane that are separated by ridges, where the elements remain intact.

\begin{figure}
  \centerline{\includegraphics[width=.48\textwidth,height=.23\textheight]{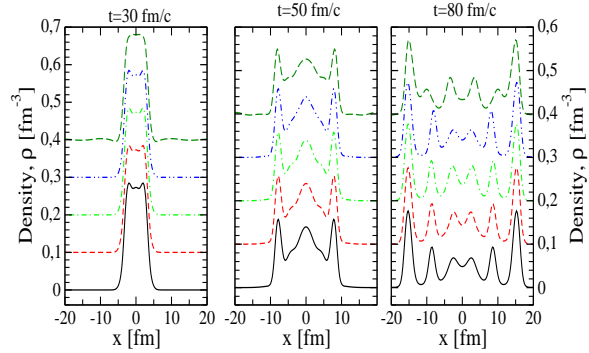}}
  \caption{Density at different stages of the evolution at the incident c.m.\ energy of 25~MeV/nucleon.  Bottom curves represent the density from the standard evolution.  The other curves, from bottom up, show the density obtained from an evolution where the matrix elements at $|x - x'| > 20$, 15, 10, and 5~fm, respectively, are suppressed.  For clarity, the results for the density of the different evolutions are staggered by 0.1~fm$^{-3}$.}
  \label{fig:suppress}
\end{figure}

\begin{figure}
  \centerline{\includegraphics[width=.35\textwidth]{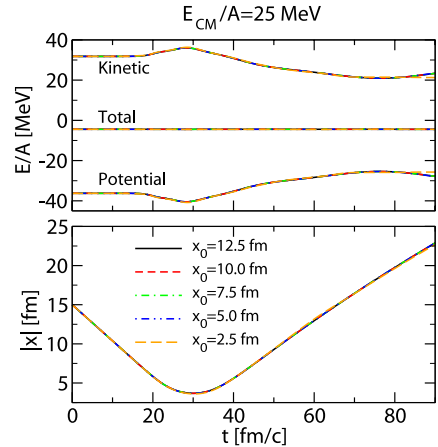}}
  \caption{Net energy components (top) and system size (bottom) for elements of density matrix suppressed at $|x - x'| > 2 x_0$, at different values of $x_0$.}
  \label{fig:energy_suppress}
\end{figure}

Figure~\ref{fig:suppress} compares the results for the density at
three characteristic times in the slab collision at $E_\text{cm} = 25
\, \text{MeV/nucleon}$. Each line corresponds to evolutions with
different retained regions off the diagonal.  Only when the matrix elements are suppressed from $|x - x'|= 5 \, \text{fm}$ on, some changes in the density, compared to the standard evolution, begin to emerge at late times.  Even these changes are subtle and the system may still reach about the same final state, just slightly later.  Less severe trimmings of the elements leave no visible signs in the evolution of density for times relevant for reaction dynamics.  Figure \ref{fig:energy_suppress} shows in addition the evolution of the system size and of the net energy components for different off-diagonal suppressions.  Only for elements eliminated at $|x - x'| > 5 \, \text{fm}$, a minute change in the energy breakdown begins to emerge at late times.

Reasoning that relative phases of amplitudes associated with different fragments would matter only if the fragments could meet again, we have further carried out calculations where we allowed the fragments to reach the boundary of the calculational region, cross to an adjacent region and collide there again.  Even in those cases, we found no perceptible difference in the evolution of the density when elements had been eliminated at distances $|x - x'| \gtrsim 10 \, \text{fm}$.  The only situation when we find some genuine impact of the far off-diagonal elements is when trying to run the evolution backwards in time.  The original mean-field evolution is time reversible and such, to a very good extent, is its numerical realization.  However, reversibility gets lost when erasing off-diagonal elements and considering longer time intervals.  Thus, e.g.\ if we suppress elements at $|x - x'| \gtrsim 10 \, \text{fm}$, while evolving the system forward in time, past the reaction, and then evolve the system backwards with element suppression, we find that the system has a difficulty reaching its initial state~\cite{rios09}.  In general, the system evolved backwards in time, with elements suppressed, acquires, in the past, some features of the state in the future.  Thus, in the energy region for fusion, the system is likely to stay fused when evolved back to the early times.  In the energy range for multifragmentation, the system may evolve back to two residues but, between them, an additional fragment may remain and the two residues may stay excited.

\section{Summary}

To sum up, we have obtained encouraging results in our study of slab collisions in one dimension.  Only a~limited range of arguments in the density matrix is of practical importance for the evolution forward in time.  The scale for that range turns out in practice to be set by the inverse of Fermi momentum.  Further, we find that an adiabatic transformation of the Hamiltonian allows for an arrival at the ground state of an interacting, albeit mean-field, Hamiltonian.  Those results bode well for the possibility of simulating central nuclear reactions, in 3 dimensions, following the method of nonequilibrium Green's functions.

\section*{Acknowledgements}

This work was partially supported by the U.S.\ National Science Foundation under Grants PHY-0555893 and PHY-0800026 and by the Project of Knowledge Innovation Program of the Chinese Academy of Sciences under Grant KJCX2.YW.W10.



\begin{thebibliography}{10}

\bibitem{Bonche1976}
P.~Bonche, S.~Koonin, and J.~W.~Negele, Phys.\ Rev.\ C 13, 1226 (1976).

\bibitem{negele82}
J.~W.~Negele, Rev.\ Mod.\ Phys.\ 54, 913 (1982).


\bibitem{Umar05}
A.~S.~Umar, V.~E.~Oberacker, Eur.\ Phys.\ J.\ A 25, s1.553 (2005).


\bibitem{bertsch88}
G.\ F.\ Bertsch and S.\ Das Gupta, Phys.\ Rep.\ 160, 189 (1988).

\bibitem{aic86}
J.~Aichelin and H.~St\"ocker, Phys.\ Lett.~B176 (1986) 14.

\bibitem{ono92}
A.~Ono, H.~Horiuchi, T.~Maruyama and A. Ohnishi,
Phys.\ Rev.\ Lett.~68 (1992) 2898.

\bibitem{kadanoff}
L. P. Kadanoff and G. Baym, {\it Quantum Statistical Mechanics}, Benjamin (New York) 1962.

\bibitem{danielewicz84}
P. Danielewicz, Ann. Phys. 152, 239 (1984); {\it ibid.} 305 (1984).

\bibitem{krieger81}
S.~J.~Krieger and M.~S.~Weiss, Phys.\ Rev.\ C 24, 928 (1981).

\bibitem{szanto81}
A.~Szanto de Toledo {\em et al.}, Phys.\ Rev.\ Lett.\ 47, 1881 (1981).

\bibitem{dan91}
P.\ Danielewicz and G.~F.~Bertsch,
Nucl.\ Phys.\ A {533}, 712 (1991).

\bibitem{dickhoff98}
W.~H.~Dickhoff, Phys.\ Rev.\ C 58, 2807 (1998).


\bibitem{nagamiya81}
S.~Nagamiya {\em et al.}, {Phys.\ Rev.\ C} {24}, 971
(1981).

\bibitem{ivanov00}
Yu.~B.~Ivanov, J.~Knoll and D.~N.~Voskresensky,
Nucl.\ Phys.\ A 672, 313 (2000).

\bibitem{leupold00}
S.~Leupold,
Nucl.\ Phys.\ A 672, 475 (2000).


\bibitem{cassing00}
W.~Cassing and S.~Juchem,
Nucl.\ Phys.\ A 665, 377 (2000).


\bibitem{kohler95}
H.~S.~K\"ohler,
Phys.\ Rev.\ C 51, 3232 (1995).

\bibitem{haug}
H.~Haug and A.-P.~Jauho, {\it Quantum Kinetics in Transport and Optics of Semiconductors}, Springer (Berlin) 1996.

\bibitem{kwong98}
 N.~H.~Kwong {\it et al.}, Phys.\ Stat.\ Sol.\ B 206, 197 (1998).

\bibitem{dahlen07}
N.~E.~Dahlen and R.~van Leeuwen, Phys.\ Rev.\ Lett.\ 98, 153004 (2007).

\bibitem{muther00}
H.~M{\"u}ther and A.~Polls, Prog.\ Part.\ Nucl.\ Phys.\ 45, 243 (2000).

\bibitem{dickhoff04}
W.~H.~Dickhoff and C.~Barbieri, Prog.\ Part.\ Nucl.\ Phys.\ 52, 377 (2004).

\bibitem{tohyama87}
M.~Tohyama, Phys.\ Rev.\ C 36, 187 (1987).

\bibitem{messiah}
A.~Messiah, {\em Quantum Mechanics}, Dover Publications (New York) 1999.

\bibitem{rios09}
 A. Rios, P. Danielewicz, B. Barker and M. Buchler, to be published.


\end{thebibliography}

\end{document}